\documentclass[twocolumn,english,superscriptaddress, showkeys]{revtex4}
\usepackage[T1]{fontenc}
\usepackage[latin9]{inputenc}
\usepackage{color}
\usepackage{babel}
\usepackage{mathtools}
\usepackage{amsmath}
\usepackage{amssymb}
\usepackage{graphicx}
\usepackage[unicode=true,pdfusetitle,
 bookmarks=true,bookmarksnumbered=false,bookmarksopen=false,
 breaklinks=true,pdfborder={0 0 0},backref=false,colorlinks=true]
 {hyperref}

\makeatletter
\@ifundefined{textcolor}{}
{%
 \definecolor{BLACK}{gray}{0}
 \definecolor{WHITE}{gray}{1}
 \definecolor{RED}{rgb}{1,0,0}
 \definecolor{GREEN}{rgb}{0,1,0}
 \definecolor{BLUE}{rgb}{0,0,1}
 \definecolor{CYAN}{cmyk}{1,0,0,0}
 \definecolor{MAGENTA}{cmyk}{0,1,0,0}
 \definecolor{YELLOW}{cmyk}{0,0,1,0}
}

\hypersetup{colorlinks=true,citecolor=blue,linkcolor=cyan,urlcolor=blue,filecolor= green}

\makeatother

\begin{document}

\title{Computing coherence vectors and correlation matrices, \\ with application
to quantum discord quantification}

\author{Jonas Maziero}

\email{jonas.maziero@ufsm.br}

\address{Departamento de F\'isica, Centro de Ci\^encias Naturais e Exatas, Universidade Federal de Santa Maria, Avenida Roraima 1000, 97105-900, Santa Maria, RS, Brazil}

\address{Instituto de F\'isica, Facultad de Ingenier\'ia, Universidad de la Rep\'ublica, J. Herrera y Reissig 565, 11300, Montevideo, Uruguay}
\begin{abstract}
ABSTRACT\\Coherence vectors and correlation matrices are important
functions frequently used in physics. The numerical calculation of
these functions directly from their definitions, which involves Kronecker
products and matrix multiplications, may seem to be a reasonable option.
Notwithstanding, as we demonstrate in this article, some algebraic
manipulations before programming can reduce considerably their computational
complexity. Besides, we provide Fortran code to generate generalized
Gell Mann matrices and to compute the optimized and unoptimized versions
of the associated Bloch's vectors and correlation matrix, in the case
of bipartite quantum systems. As a code test and application example,
we consider the calculation of Hilbert-Schmidt quantum discords.
\end{abstract}


\maketitle

\section{Introduction}

Correlation functions are fundamental objects for statistical analysis,
and are thus ubiquitous in most kinds of scientific inquires and their
applications \cite{Jaynes,Devore}. In physics, correlation functions
have an important role for research in areas such as quantum optics
and open systems \cite{Knight_book,Petruccione_book}, phase transitions
and condensed matter physics \cite{Sachdev_book,Mezzadri_AMP}, and
quantum field theory and nuclear and particle physics \cite{Mozo_QFT}.
Another area in which correlation functions are omnipresent is quantum
information science (QIS), an interdisciplinary field that extends
the applicabilities of the classical theories of information, computation,
and computational complexity \cite{Nielsen=000026Chuang,Preskill,Wilde}.

Investigations about the quantum correlations in physical systems
have been one of the main catalyzers for developments in QIS \cite{Leuchs_RMP,Horodecki_RMP,Jost_AMP,Modi_RMP,Lucas_RevD}.
There are several guises of quantum correlations, and quantum discord
stands among the most promising quantum resources for fueling the
quantum advantage \cite{Lucas_PTRSA,Brito_AMP,Streltsov_book,Datta_DQC1,Winter_DSMerg,Pirandola_DCryp,Retamal_DSDisc,Piani_DDHid,Vedral_RSP,Adesso_metro,Piani_DEComm}.
When computing or witnessing quantum discord, or other kinds of correlation
or quantumness quantifiers, we are frequently faced with the need
for calculating coherence vectors and correlation matrices \cite{Maziero_W1,Maziero_W2,Sarandy_2sc_exp,Adesso_OD,Maziero_W3,Bose_AMP,LoFranco1,LoFranco2,LoFranco3,LoFranco4,LoFranco5,LoFranco6,LoFranco7,LoFranco8,LoFranco9}.
And it is the main aim of this article to provide formulas for these
functions that are amenable for more efficient numerical calculations
when compared with the direct implementation of their definitions.

In order to define coherence vectors and correlation matrices, let
us consider a composite bipartite system with Hilbert space $\mathcal{H}_{ab}=\mathcal{H}_{a}\otimes\mathcal{H}_{b}$.
Hereafter the corresponding dimensions are denoted by $d_{s}=\dim\mathcal{H}_{s}$
for $s=ab,a,b$. In addition, let $\Gamma_{j}^{s}$, with
\begin{equation}
\mathrm{Tr}(\Gamma_{j}^{s})=0\mbox{ and }\mathrm{Tr}(\Gamma_{j}^{s}\Gamma_{k}^{s})=2\delta_{jk},
\end{equation}
be a basis for the special unitary group $SU(d_{s})$. Any density
operator describing the state of the system $\mathcal{H}_{ab}$ can
be written in the local basis $\Gamma_{j}^{a}\otimes\Gamma_{k}^{b}$
as follows:
\begin{eqnarray}
\rho & = & \frac{1}{d_{ab}}\left(\mathbb{I}_{a}\otimes\mathbb{I}_{b}+\sum_{j=1}^{d_{a}^{2}-1}a_{j}\Gamma_{j}^{a}\otimes\mathbb{I}_{b}+\mathbb{I}_{a}\otimes\sum_{k=1}^{d_{b}^{2}-1}b_{k}\Gamma_{k}^{b}\right.\nonumber \\
 &  & \hspace{1em}\hspace{1em}\hspace{1em}\left.+\sum_{j=1}^{d_{a}^{2}-1}\sum_{k=1}^{d_{b}^{2}-1}c_{j,k}\Gamma_{j}^{a}\otimes\Gamma_{k}^{b}\right),\label{eq:rho_Bloch}
\end{eqnarray}
where
\begin{equation}
j=1,\cdots,d_{a}^{2}-1\mbox{ and }k=1,\cdots,d_{b}^{2}-1,
\end{equation}
and $\mathbb{I}_{s}$ is the identity operator in $\mathcal{H}_{s}$.
One can readily verify that the components of the coherence (or Bloch's)
vectors $\mathbf{a}=(a_{1},\cdots,a_{d_{a}^{2}-1})$ and $\mathbf{b}=(b_{1},\cdots,b_{d_{b}^{2}-1})$
and of the correlation matrix $C=(c_{j,k})$ are given by: 
\begin{eqnarray}
a_{j} & = & 2^{-1}d_{a}\mathrm{Tr}(\Gamma_{j}^{a}\otimes\mathbb{I}_{b}\rho),\\
b_{k} & = & 2^{-1}d_{b}\mathrm{Tr}(\mathbb{I}_{a}\otimes\Gamma_{k}^{b}\rho),\\
c_{j,k} & = & 2^{-2}d_{ab}\mathrm{Tr}(\Gamma_{j}^{a}\otimes\Gamma_{k}^{b}\rho).
\end{eqnarray}
It is worthwhile mentioning that the mean value of \emph{any observable}
in $\mathcal{H}_{s}$, for $s=a,b,ab$, can be obtained using these
quantities. 

In \href{https://github.com/jonasmaziero/LibForQ.git}{https://github.com/jonasmaziero/LibForQ.git},
we provide Fortran code to compute the coherence vectors, correlation
matrices, and quantum discord quantifiers we deal with here. Besides
these functions, there are other tools therein that may be of interest
to the reader. The instructions on how to use the software are provided
in the readme file. Related to the content of this section, the subroutine
\texttt{bloch\_vector\_gellmann\_unopt($d_{s}$, $\rho_{s}$, $\mathbf{s}$)}
returns the coherence vectors $\mathbf{a}$ or $\mathbf{b}$ and the
subroutine \texttt{corrmat\_gellmann\_unopt($d_{a}$, $d_{b}$, $\rho$,
$C$)} computes the correlation matrix $C$. Now, let us notice that
if calculated directly from the equations above, for $d_{a},d_{b}\gg1$,
the computational complexity (CC) to obtain the coherence vectors
$\mathbf{a}$ and $\mathbf{b}$ or the correlation matrix $C$ is:
\begin{equation}
CC(\mathbf{a})=CC(\mathbf{b})=CC(C)\approx\mathcal{O}(d_{a}^{6}d_{b}^{6}).
\end{equation}

The remainder of this article is structured as follows. In Sec. \ref{coh_vec},
we obtain formulas for $\mathbf{a}$, $\mathbf{b}$, and $C$ that
are amenable for more efficient numerical computations. In Sec. \ref{discord}
we test these formulas by applying them in the calculation of Hilbert-Schmidt
quantum discords. In Sec. \ref{conclusion} we make some final remarks
about the usefulness and possible applications of the results reported
here.

\section{Computing coherence vectors and correlation matrices}

\label{coh_vec}

The partial trace function \cite{Maziero_PTr} can be used in order
to obtain the reduced states $\rho_{a}=\mathrm{Tr}_{b}(\rho)$ and
$\rho_{b}=\mathrm{Tr}_{a}(\rho)$ and to write the components of the
Bloch vectors in the form:
\begin{eqnarray}
a_{j} & = & 2^{-1}d_{a}\mathrm{Tr}(\Gamma_{j}^{a}\rho_{a}),\\
b_{k} & = & 2^{-1}d_{b}\mathrm{Tr}(\Gamma_{k}^{b}\rho_{b}).
\end{eqnarray}
Thus, when computing the coherence vectors of the parties $a$ and
$b$, we shall have to solve a similar problem; so let's consider
it separately. That is to say, we shall regard a generic density operator
written as
\begin{equation}
\rho_{s}=\frac{1}{d_{s}}\left(\mathbb{I}_{s}+\sum_{j=1}^{d^{2}-1}s_{j}\Gamma_{j}^{s}\right),
\end{equation}
where $s_{j}=2^{-1}d_{s}\mathrm{Tr}(\rho_{s}\Gamma_{j}^{s}).$ 

Now, and for the remainder of this article, we assume that the matrix
elements of regarded density operator $\rho$ in the standard computational
basis are given. We want to compute the \emph{Bloch's vector} \cite{Petruccione_param}:
$\mathbf{s}=(s_{1},\cdots,s_{d_{s}^{2}-1}).$ For the sake of doing
that, a particular basis $\Gamma_{j}^{s}$ must be chosen. Here we
pick the generalized Gell Mann matrices, which are presented below
in three groups \cite{Bertlmann}:
\begin{eqnarray}
 &  & \Gamma_{j}^{s(1)}=\sqrt{\frac{2}{j(j+1)}}\sum_{k=1}^{j+1}(-j)^{\delta_{k,j+1}}|k\rangle\langle k|,\label{eq:SU1}\\
 &  & \mbox{ }\mbox{ }\mbox{ }\mbox{ for }j=1,\cdots,d_{s}-1,\nonumber \\
 &  & \Gamma_{(k,l)}^{s(2)}=|k\rangle\langle l|+|l\rangle\langle k|,\mbox{ for }1\leq k<l\leq d_{s},\label{eq:SU2}\\
 &  & \Gamma_{(k,l)}^{s(3)}=-i(|k\rangle\langle l|-|l\rangle\langle k|),\mbox{ for }1\leq k<l\leq d_{s},\label{eq:SU3}
\end{eqnarray}
which are named the diagonal, symmetric, and antisymmetric group,
respectively. The last two groups possess $d_{s}(d_{s}-1)/2$ generators
each. Any one of these matrices can be obtained by calling the subroutine
\texttt{gellmann($d_{s}$, $g$, $k$, $l$, $\Gamma_{(k,l)}^{s(g)}$)}.
For the first group, $g=1$, we make $j=k$ and, in this case, one
can set $l$ to any integer.

It is straightforward seeing that, for the generators above, the corresponding
components of the Bloch's vector can expressed directly in terms of
the matrix elements of the density operator $\rho_{s}$ as follows:
\begin{eqnarray}
 &  & s_{j}^{(1)}=\frac{d_{s}}{\sqrt{2j(j+1)}}\sum_{k=1}^{j+1}(-j)^{\delta_{k,j+1}}\langle k|\rho_{s}|k\rangle,\\
 &  & \mbox{ }\mbox{ }\mbox{ }\mbox{for }j=1,\cdots,d_{s}-1,\nonumber \\
 &  & s_{(k,l)}^{(2)}=d_{s}\mathrm{Re}\langle l|\rho_{s}|k\rangle,\mbox{ for }1\leq k<l\leq d_{s},\\
 &  & s_{(k,l)}^{(2)}=d_{s}\mathrm{Im}\langle l|\rho_{s}|k\rangle,\mbox{ for }1\leq k<l\leq d_{s}.
\end{eqnarray}

These expressions were implemented in the Fortran subroutine \texttt{bloch\_vector\_gellmann($d_{s}$,
$\rho_{s}$, $\mathbf{s}$)}. With this subroutine, and the partial
trace function \cite{Maziero_PTr}, we can compute the coherence vectors
$\mathbf{a}$ and $\mathbf{b}$. 

We observe that after these simple algebraic manipulations the computational
complexity of the Bloch's vector turns out to be basically the CC
for the partial trace function. Hence, from Ref. \cite{Maziero_PTr}
we have that for $d_{a},d_{b}\gg1$,
\begin{equation}
CC(\mathbf{a})\approx\mathcal{O}(d_{a}^{2}d_{b})\mbox{ and }CC(\mathbf{b})\approx\mathcal{O}(d_{a}d_{b}^{2}).
\end{equation}

One detail we should keep in mind, when making use of the codes linked
to this article, is the convention we apply for the indexes of the
components of $\mathbf{s}$. For the first group of generators, $\Gamma_{j}^{s(1)}$,
naturally, $j=1,\cdots,d_{s}-1$. We continue with the second group
of generators, $\Gamma_{j}^{s(2)}=\Gamma_{(k,l)}^{s(2)}$, by setting
$j_{(k,l)=(1,2)}=d_{s}-1+1=d_{s}$, $j_{(k,l)=(1,3)}=d_{s}+1$, $\cdots$,
$j_{(k,l)=(1,d_{s})}=2(d_{s}-1)$, $j_{(k,l)=(2,3)}=2(d_{s}-1)+1$,
$\cdots$. The same convention is used for the third group of generators,
$\Gamma_{j}^{s(3)}=\Gamma_{(k,l)}^{s(3)}$, but here we begin with
$j_{(k,l)=(1,2)}=d_{s}-1+2^{-1}d_{s}(d_{s}-1)+1=d_{s}+2^{-1}d_{s}(d_{s}-1).$

Next we address the computation of the \emph{correlation matrix} $C=(c_{j,k})$,
which is a $(d_{a}^{2}-1)\mathrm{x}(d_{b}^{2}-1)$ matrix that we
write in the form:
\begin{equation}
C=\begin{bmatrix}C^{(1,1)} & C^{(1,2)} & C^{(1,3)}\\
C^{(2,1)} & C^{(2,2)} & C^{(2,3)}\\
C^{(3,1)} & C^{(3,2)} & C^{(3,3)}
\end{bmatrix},\label{eq:corrmat}
\end{equation}
with the sub-matrices given as shown below. For convenience, we define
the auxiliary variables:
\begin{equation}
\iota\coloneqq\sqrt{\frac{2}{j(j+1)}}\mbox{, }\kappa\coloneqq\sqrt{\frac{2}{k(k+1)}}\mbox{, and }\varsigma\coloneqq\frac{d_{a}d_{b}}{4}.
\end{equation}

The matrix elements of $C^{(1,1)}$, whose dimension is $(d_{a}-1)\mathrm{x}(d_{b}-1)$,
correspond to the diagonal generators for $a$ and diagonal generators
for $b$:
\begin{eqnarray}
 &  & c_{j,k}^{(1,1)}=\varsigma\mathrm{Tr}(\Gamma_{j}^{a(1)}\otimes\Gamma_{k}^{b(1)}\rho)\nonumber \\
 & = & \varsigma\iota\kappa\mathrm{Tr}(\sum_{m=1}^{j+1}(-j)^{\delta_{m,j+1}}|m\rangle\langle m|)\otimes(\sum_{p=1}^{k+1}(-k)^{\delta_{p,k+1}}|p\rangle\langle p|)\rho\nonumber \\
 & = & \varsigma\iota\kappa\sum_{m=1}^{j+1}\sum_{p=1}^{k+1}(-j)^{\delta_{m,j+1}}(-k)^{\delta_{p,k+1}}\mathrm{Tr}(|m\rangle\langle m|\otimes|p\rangle\langle p|\rho)\nonumber \\
 & = & \varsigma\iota\kappa\sum_{m=1}^{j+1}\sum_{p=1}^{k+1}(-j)^{\delta_{m,j+1}}(-k)^{\delta_{p,k+1}}\langle mp|\rho|mp\rangle.\label{eq:c11}
\end{eqnarray}

The matrix elements of $C^{(1,2)}$, whose dimension is $(d_{a}-1)\mathrm{x}2^{-1}d_{b}(d_{b}-1)$,
correspond to the diagonal generators for $a$ and symmetric generators
for $b$:
\begin{eqnarray}
 &  & c_{j,k}^{(1,2)}=\varsigma\mathrm{Tr}(\Gamma_{j}^{a(1)}\otimes\Gamma_{k}^{b(2)}\rho)=\varsigma\mathrm{Tr}(\Gamma_{j}^{a(1)}\otimes\Gamma_{(p,q)}^{b(2)}\rho)\nonumber \\
 & = & \varsigma\mathrm{Tr}\left(\iota(\sum_{m=1}^{j+1}(-j)^{\delta_{m,j+1}}|m\rangle\langle m|)\otimes(|p\rangle\langle q|+|q\rangle\langle p|)\rho\right)\nonumber \\
 & = & \varsigma\iota\sum_{m=1}^{j+1}(-j)^{\delta_{m,j+1}}\left(\mathrm{Tr}(|mp\rangle\langle mq|\rho)+\mathrm{Tr}(|mq\rangle\langle mp|\rho)\right)\nonumber \\
 & = & \varsigma\iota\sum_{m=1}^{j+1}(-j)^{\delta_{m,j+1}}\left(\langle mq|\rho|mp\rangle+\langle mq|\rho|mp\rangle^{*}\right)\nonumber \\
 & = & 2\varsigma\iota\sum_{m=1}^{j+1}(-j)^{\delta_{m,j+1}}\mathrm{Re}\langle mq|\rho|mp\rangle.\label{eq:c12}
\end{eqnarray}

The matrix elements of $C^{(1,3)}$, whose dimension is $(d_{a}-1)\mathrm{x}2^{-1}d_{b}(d_{b}-1)$,
correspond to the diagonal generators for $a$ and antisymmetric generators
for $b$:
\begin{eqnarray}
 &  & c_{j,k}^{(1,3)}=\varsigma\mathrm{Tr}(\Gamma_{j}^{a(1)}\otimes\Gamma_{k}^{b(3)}\rho)=\varsigma\mathrm{Tr}(\Gamma_{j}^{a(1)}\otimes\Gamma_{(p,q)}^{b(3)}\rho)\nonumber \\
 & = & -i\varsigma\iota\mathrm{Tr}\left((\sum_{m=1}^{j+1}(-j)^{\delta_{m,j+1}}|m\rangle\langle m|)\otimes(|p\rangle\langle q|-|q\rangle\langle p|)\rho\right)\nonumber \\
 & = & -i\varsigma\iota\sum_{m=1}^{j+1}(-j)^{\delta_{m,j+1}}\left(\mathrm{Tr}(|mp\rangle\langle mq|\rho)-\mathrm{Tr}(|mq\rangle\langle mp|\rho)\right)\nonumber \\
 & = & -i\varsigma\iota\sum_{m=1}^{j+1}(-j)^{\delta_{m,j+1}}\left(\langle mq|\rho|mp\rangle-\langle mq|\rho|mp\rangle^{*}\right)\nonumber \\
 & = & 2\varsigma\iota\sum_{m=1}^{j+1}(-j)^{\delta_{m,j+1}}\mathrm{Im}\langle mq|\rho|mp\rangle.\label{eq:c13}
\end{eqnarray}

The matrix elements of $C^{(2,1)}$, whose dimension is $2^{-1}d_{a}(d_{a}-1)\mathrm{x}(d_{b}-1)$,
correspond to the symmetric generators for $a$ and diagonal generators
for $b$:
\begin{eqnarray}
 &  & c_{j,k}^{(2,1)}=\varsigma\mathrm{Tr}(\Gamma_{j}^{a(2)}\otimes\Gamma_{k}^{b(1)}\rho)=\varsigma\mathrm{Tr}(\Gamma_{(m,n)}^{a(2)}\otimes\Gamma_{k}^{b(1)}\rho)\nonumber \\
 & = & \varsigma\mathrm{Tr}\left((|m\rangle\langle n|+|n\rangle\langle m|)\otimes\kappa(\sum_{p=1}^{k+1}(-k)^{\delta_{p,k+1}}|p\rangle\langle p|)\rho\right)\nonumber \\
 & = & \varsigma\kappa\sum_{p=1}^{k+1}(-k)^{\delta_{p,k+1}}\left(\mathrm{Tr}(|mp\rangle\langle np|\rho)+\mathrm{Tr}(|np\rangle\langle mp|\rho)\right)\nonumber \\
 & = & \varsigma\kappa\sum_{p=1}^{k+1}(-k)^{\delta_{p,k+1}}\left(\langle np|\rho|mp\rangle+\langle np|\rho|mp\rangle^{*}\right)\nonumber \\
 & = & 2\varsigma\kappa\sum_{p=1}^{k+1}(-k)^{\delta_{p,k+1}}\mathrm{Re}\langle np|\rho|mp\rangle.\label{eq:c21}
\end{eqnarray}

The matrix elements of $C^{(2,2)}$, whose dimension is $2^{-1}d_{a}(d_{a}-1)\mathrm{x}2^{-1}d_{b}(d_{b}-1)$,
correspond to the symmetric generators for $a$ and symmetric generators
for $b$:
\begin{eqnarray}
 &  & c_{j,k}^{(2,2)}=\varsigma\mathrm{Tr}(\Gamma_{j}^{a(2)}\otimes\Gamma_{k}^{b(2)}\rho)=\varsigma\mathrm{Tr}(\Gamma_{(m,n)}^{a(2)}\otimes\Gamma_{(p,q)}^{b(2)}\rho)\nonumber \\
 &  & =\varsigma\mathrm{Tr}(\left(|m\rangle\langle n|+|n\rangle\langle m|\right)\otimes\left(|p\rangle\langle q|+|q\rangle\langle p|\right)\rho)\nonumber \\
 &  & =\varsigma\left(\langle nq|\rho|mp\rangle+\langle mq|\rho|np\rangle+\langle np|\rho|mq\rangle+\langle mp|\rho|nq\rangle\right)\nonumber \\
 &  & =2\varsigma\left(\mathrm{Re}\langle nq|\rho|mp\rangle+\mathrm{Re}\langle np|\rho|mq\rangle\right).\label{eq:c22}
\end{eqnarray}

The matrix elements of $C^{(2,3)}$, whose dimension is $2^{-1}d_{a}(d_{a}-1)\mathrm{x}2^{-1}d_{b}(d_{b}-1)$,
correspond to the symmetric generators for $a$ and antisymmetric
generators for $b$:
\begin{eqnarray}
 &  & c_{j,k}^{(2,3)}=\varsigma\mathrm{Tr}(\Gamma_{j}^{a(2)}\otimes\Gamma_{k}^{b(3)}\rho)=\varsigma\mathrm{Tr}(\Gamma_{(m,n)}^{a(2)}\otimes\Gamma_{(p,q)}^{b(3)}\rho)\nonumber \\
 &  & =\varsigma\mathrm{Tr}(\left(|m\rangle\langle n|+|n\rangle\langle m|\right)\otimes(-i)\left(|p\rangle\langle q|-|q\rangle\langle p|\right)\rho)\nonumber \\
 &  & =-i\varsigma\left(\langle nq|\rho|mp\rangle+\langle mq|\rho|np\rangle-\langle np|\rho|mq\rangle-\langle mp|\rho|nq\rangle\right)\nonumber \\
 &  & =2\varsigma\left(\mathrm{Im}\langle nq|\rho|mp\rangle-\mathrm{Im}\langle np|\rho|mq\rangle\right).\label{eq:c23}
\end{eqnarray}

The matrix elements of $C^{(3,1)}$, whose dimension is $2^{-1}d_{a}(d_{a}-1)\mathrm{x}(d_{b}-1)$,
correspond to the antisymmetric generators for $a$ and diagonal generators
for $b$:
\begin{eqnarray}
 &  & c_{j,k}^{(3,1)}=\varsigma\mathrm{Tr}(\Gamma_{j}^{a(3)}\otimes\Gamma_{k}^{b(1)}\rho)=\mathrm{Tr}(\Gamma_{(m,n)}^{a(3)}\otimes\Gamma_{k}^{b(1)}\rho)\nonumber \\
 &  & =\varsigma\mathrm{Tr}\left(-i(|m\rangle\langle n|-|n\rangle\langle m|)\otimes\kappa\sum_{p=1}^{k+1}(-k)^{\delta_{p,k+1}}|p\rangle\langle p|\rho\right)\nonumber \\
 &  & =-i\varsigma\kappa\sum_{p=1}^{k+1}(-k)^{\delta_{p,k+1}}\left(\langle np|\rho|mp\rangle-\langle mp|\rho|np\rangle\right)\nonumber \\
 &  & =2\varsigma\kappa\sum_{p=1}^{k+1}(-k)^{\delta_{p,k+1}}\mathrm{Im}\langle np|\rho|mp\rangle.\label{eq:c31}
\end{eqnarray}

The matrix elements of $C^{(3,2)}$, whose dimension is $2^{-1}d_{a}(d_{a}-1)\mathrm{x}2^{-1}d_{b}(d_{b}-1)$,
correspond to the antisymmetric generators for $a$ and symmetric
generators for $b$:
\begin{eqnarray}
 &  & c_{j,k}^{(3,2)}=\varsigma\mathrm{Tr}(\Gamma_{j}^{a(3)}\otimes\Gamma_{k}^{b(2)}\rho)=\varsigma\mathrm{Tr}(\Gamma_{(m,n)}^{a(3)}\otimes\Gamma_{(p,q)}^{b(2)}\rho)\nonumber \\
 &  & =\varsigma\mathrm{Tr}\left((-i)\left(|m\rangle\langle n|-|n\rangle\langle m|\right)\otimes\left(|p\rangle\langle q|+|q\rangle\langle p|\right)\rho\right)\nonumber \\
 &  & =-i\varsigma\left(\langle nq|\rho|mp\rangle-\langle mp|\rho|nq\rangle+\langle np|\rho|mq\rangle-\langle mq|\rho|np\rangle\right)\nonumber \\
 &  & =2\varsigma\left(\mathrm{Im}\langle nq|\rho|mp\rangle+\mathrm{Im}\langle np|\rho|mq\rangle\right).\label{eq:c32}
\end{eqnarray}

The matrix elements of $C^{(3,3)}$, whose dimension is $2^{-1}d_{a}(d_{a}-1)\mathrm{x}2^{-1}d_{b}(d_{b}-1)$,
correspond to antisymmetric generators for $a$ and antisymmetric
generators for $b$:
\begin{eqnarray}
 &  & c_{j,k}^{(3,3)}=\varsigma\mathrm{Tr}(\Gamma_{j}^{a(3)}\otimes\Gamma_{k}^{b(3)}\rho)=\varsigma\mathrm{Tr}(\Gamma_{(m,n)}^{a(3)}\otimes\Gamma_{(p,q)}^{b(3)}\rho)\nonumber \\
 &  & =\varsigma\mathrm{Tr}\left(-i\left(|m\rangle\langle n|-|n\rangle\langle m|\right)\otimes(-i)\left(|p\rangle\langle q|-|q\rangle\langle p|\right)\rho\right)\nonumber \\
 &  & =-\varsigma\left(\langle nq|\rho|mp\rangle+\langle mp|\rho|nq\rangle-\langle np|\rho|mq\rangle-\langle mq|\rho|np\rangle\right)\nonumber \\
 &  & =2\varsigma\left(\mathrm{Re}\langle np|\rho|mq\rangle-\mathrm{Re}\langle nq|\rho|mp\rangle\right).\label{eq:c33}
\end{eqnarray}

We remark that when implementing these expressions numerically, for
the sake of mapping the local to the global computational basis, we
utilize, e.g.,
\begin{equation}
|np\rangle\equiv|(n-1)d_{b}+p\rangle.
\end{equation}

The subroutine \texttt{corrmat\_gellmann($d_{a}$, $d_{b}$, $\rho$,
$C$)} returns the correlation matrix $C=(c_{j,k})$, as written in
Eq. (\ref{eq:corrmat}), associated with the bipartite density operator
$\rho$ and computed using the Gell Mann basis, as described in this
section. The convention for the indexes of the matrix elements $c_{j,k}$
is defined in the same way as for the coherence vectors. The computational
complexity for $C$, computed via the optimized expressions obtained
in this section, is, for $d_{a},d_{b}\gg1$,
\begin{equation}
CC(C)\approx\mathcal{O}(d_{a}^{2}d_{b}^{2}).
\end{equation}
By generating some random density matrices \cite{Maziero_FICT}, we
checked that the expressions and the corresponding code for the unoptimized
and optimized versions of $\mathbf{a}$, $\mathbf{b}$, and $C$ agree.
Additional tests shall be presented in the next section, where we
calculate some quantum discord quantifiers.

\section{Computing Hilbert-Schmidt quantum discords}

\label{discord}

The calculation of quantum discord functions (QD) usually involves
hard optimization problems \cite{Huang_DiscCC,Sen_DiscCC}. In the
last few years, a large amount of effort have been dedicated towards
computing QD analytically, with success being obtained mostly for
low dimensional quantum systems \cite{Luo_DiscAn1,Luo_DiscAn2,Alber_DiscAnal_XS,Maziero_DiscAn,Adesso_DAnal,Oh_DAnal,Joag_DhsAnal,Giovannetti_AnalDtr,Orzag_DAnal,Zahir_DAnal,Seddik_DAnal,Bordone_DAnal,Fan_DAnal,Fei_DAnal,Wang_DAnal,Zhang_DAnal,Sarandy_Dtr_Anal,Huang_2}.
Although not meeting all the required properties for a \emph{bona
fide} QD quantifier \cite{Modi_DCriteria}, the Hilbert-Schmidt discord
(HSD) \cite{Brukner_HSD},
\begin{equation}
D_{hs}^{a}(\rho)=\min_{\rho_{cq}}||\rho-\rho_{cq}||_{2}^{2},
\end{equation}
is drawing much attention due to its amenability for analytical computations,
when compared with most other QD measures. In the last equation, the
minimization is performed over the classical-quantum states
\begin{equation}
\rho_{cq}=\sum_{j}p_{j}|a_{j}\rangle\langle a_{j}|\otimes\rho_{j}^{b},
\end{equation}
with $p_{j}$ being a probability distribution, $|a_{j}\rangle$ an
orthonormal basis for $\mathcal{H}_{a}$, $\rho_{j}^{b}$ generic
density operators defined in $\mathcal{H}_{b}$, and $||O||_{2}\coloneqq\sqrt{\mathrm{Tr}(O^{\dagger}O)}$
is the Hilbert-Schmidt norm of the linear operator $O$, with $O^{\dagger}$
being the transpose conjugate of $O$.

In this article, as a basic test for the Fortran code provided to
obtain coherence vectors and correlation matrices, we shall compute
the following lower bound for the HSD \cite{Luo_HSD}:
\begin{equation}
D_{hs}^{a}(\rho)=\sum_{j=d_{a}}^{d_{a}^{2}-1}\lambda_{j}^{a},
\end{equation}
where $\lambda_{j}^{a}$ are the eigenvalues, sorted in non-increasing
order, of the $(d_{a}^{2}-1)\mathrm{x}(d_{a}^{2}-1)$ matrix:
\begin{equation}
\Xi_{a}=\frac{2}{d_{a}^{2}d_{b}}\left(\mathbf{a}\mathbf{a}^{t}+\frac{2}{d_{b}}CC^{t}\right).
\end{equation}
In the equation above $t$ stands for the transpose of a vector or
matrix. We observe that the other version of the HSD, $D_{hs}^{b}$,
can be obtained from the equations above simply by exchanging $a$
and $b$ and using $C^{t}C$ instead of $CC^{t}$.

It is interesting regarding that, as was proved in Ref. \cite{Akhtarshenas1},
a bipartite state $\rho$, with polarization vectors $\mathbf{a}$
and $\mathbf{b}$ and correlation matrix $C$, is classical-quantum
if and only if there exists a $(d_{a}-1)$-dimensional projector $\Pi_{a}$
in the space $\mathbb{R}^{d_{a}^{2}-1}$ such that: 
\begin{equation}
\Pi_{a}\mathbf{a}=\mathbf{a}\mbox{ and }\Pi_{a}C=C,
\end{equation}
Based on this fact, an ameliorated version for the Hilbert-Schmidt
quantum discord (AHSD) was proposed \cite{Akhtarshenas1}: 
\begin{equation}
D_{hsa}^{a}(\rho)\coloneqq\min_{\Pi_{a}}||\Upsilon_{a}-\Pi_{a}\Upsilon_{a}||_{2}^{2},
\end{equation}
with the matrix $\Upsilon_{a}$ defined as 
\begin{equation}
\Upsilon_{a}\coloneqq\sqrt{\frac{2}{d_{a}^{2}d_{b}}}\left(f(b)\mathbf{a}\hspace{1em}g(b)\sqrt{\frac{2}{d_{b}}}C\right),
\end{equation}
where $f$ and $g$ are arbitrary functions of $b\equiv||\mathbf{b}||_{2}$.
Then, by setting $f(b)=g(b)=P(\rho_{b})$ and using the purity,
\begin{eqnarray}
P(\rho_{b}) & \coloneqq & \mathrm{Tr}(\rho_{b}^{2})=\sum_{j,k}|\rho_{j,k}^{b}|^{2},
\end{eqnarray}
to address the problem of non-contractivity of the Hilbert-Schmidt
distance, the following analytical formula was presented \cite{Akhtarshenas1}:
\begin{equation}
D_{hsa}^{a}(\rho)=\frac{1}{P(\rho_{b})}\sum_{j=d_{a}}^{d_{a}^{2}-1}\lambda_{j}^{a}=\frac{D_{hs}^{a}(\rho)}{P(\rho_{b})}.\label{eq:AHSD}
\end{equation}

\begin{figure}[h]
\begin{centering}
\includegraphics[scale=0.42]{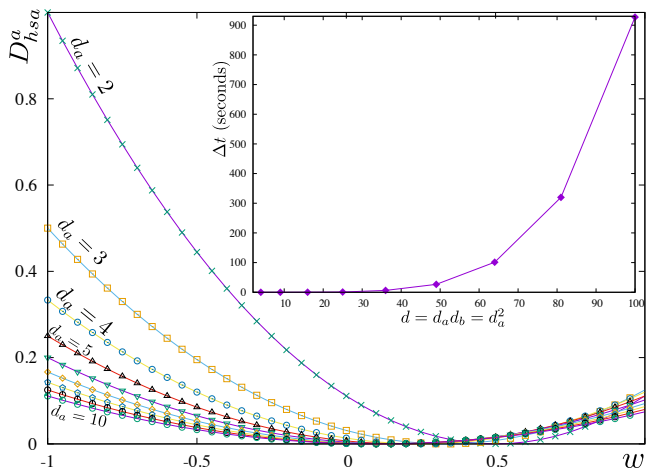}
\par\end{centering}

\caption{The points are the values of the ameliorated Hilbert-Schmidt quantum
discord of Werner states computed numerically using Eq. (\ref{eq:AHSD}).
The lines are the corresponding values of the AHSD plotted via the
analytical formula: $D_{hsa}^{a}(\rho^{w})=d_{a}D_{hs}^{a}(\rho^{w})=(d_{a}w-1)^{2}/((d_{a}-1)(d_{a}+1)^{2})$.
Due to the symmetry of $\rho^{w}$, here $D_{hsa}^{a}(\rho^{w})=D_{hsa}^{b}(\rho^{w}).$
In the inset is shown the difference between the times taken by the
two methods to compute the AHSD for a fixed value of $d_{a}$. We
see clearly here that our optimized algorithm gives a exponential
speedup against the brute force calculation of the Bloch's vectors
and correlation matrix. }

\label{werner}
\end{figure}

Thus both discord quantifiers $D_{hs}^{a}$ and $D_{hsa}^{a}$ are,
in the end of the day, obtained from the eigenvalues $\lambda_{j}^{a}$.
And the computation of these eigenvalues requires the knowledge of
the coherence vector $\mathbf{a}$ (or $\mathbf{b}$) and of the correlation
matrix $C$. These QD measures were implemented in the Fortran functions
\texttt{discord\_hs(ssys, $d_{a}$, $d_{b}$, $\rho$)} and \texttt{discord\_hsa(ssys,
$d_{a}$, $d_{b}$, $\rho$)}, where \texttt{ssys = `s'}, with $s=a,b$,
specifies which version of the quantum discord is to be computed.
As an example, let us use the formulas provided in this article and
the associated code to compute the HSD and AHSD of Werner states in
$\mathcal{H}_{a}\otimes\mathcal{H}_{b}$ (with $d_{a}=d_{b})$:
\begin{equation}
\rho^{w}=\frac{d_{a}-w}{d_{a}(d_{a}^{2}-1)}\mathbb{I}_{a}\otimes\mathbb{I}_{b}+\frac{d_{a}w-1}{d_{a}(d_{a}^{2}-1)}F,
\end{equation}
where $w\in[-1,1]$ and $F=\sum_{j,k=1}^{d_{a}}|jk\rangle\langle kj|.$
The reduced states of $\rho^{w}$ are $\mathbb{I}_{s}/d_{s}$, whose
purity is $P(\rho_{s})=1/d_{s}$. The results for the HSD and AHSD
of $\rho^{w}$ are presented in Fig. \ref{werner}.

\section{Concluding Remarks}

\label{conclusion}

In this article, we addressed the problem of computing coherence vectors
and correlations matrices. We obtained formulas for these functions
that make possible a considerably more efficient numerical implementation
when compared with the direct use of their definitions. We provided
Fortran code to calculate all the quantities regarded in this paper.
As a test for our formulas and code, we computed Hilbert-Schmidt quantum
discords of Werner states. It is important observing that, although
our focus here was in quantum information science, the tools provided
can find application in other areas, such as e.g. in the calculation
of order parameters and correlations functions for the study of phase
transitions in discrete quantum or classical systems.

\section*{Conflict of Interests}

The author declares that the funding mentioned in the Acknowledgments
section do not lead to any conflict of interest. Additionally, the
author declares that there is no conflict of interest regarding the
publication of this manuscript.
\begin{acknowledgments}
This work was supported by the Brazilian funding agencies: Conselho Nacional de Desenvolvimento Cient\'ifico e Tecnol\'ogico (CNPq), processes 441875/2014-9 and 303496/2014-2, Instituto Nacional de Ci\^encia e Tecnologia de Informa\c{c}\~ao Qu\^antica (INCT-IQ), process 2008/57856-6, and Coordena\c{c}\~ao de Desenvolvimento de Pessoal de N\'{i}vel Superior (CAPES), process 6531/2014-08. I gratefully acknowledges the hospitality of the Physics Institute and Laser Spectroscopy Group at the Universidad de la Rep\'{u}blica, Uruguay.\end{acknowledgments}


\begin{thebibliography}{10}
\bibitem{Jaynes} E. T. Jaynes, \emph{Probability Theory, The Logic
of Science} (Cambridge University Press, New York, 2003).

\bibitem{Devore} M. A. Carlton and J. L. Devore, \emph{Probability
with Applications in Engineering, Science, and Technology} (Springer,
New York, 2014).

\bibitem{Knight_book} C. Gerry and P. Knight, \emph{Introductory
Quantum Optics} (Cambridge University Press, New York, 2005).

\bibitem{Petruccione_book} H.-P. Breuer and F. Petruccione, \emph{The
Theory of Open Quantum Systems} (Oxford University Press, New York,
2002).

\bibitem{Sachdev_book} S. Sachdev, \emph{Quantum Phase Transitions}
(Cambridge University Press, New York, 2011).

\bibitem{Mezzadri_AMP} J. Hutchinson, J. P. Keating, and F. Mezzadri,
On relations between one-dimensional quantum and two-dimensional classical
spin systems, \href{http://dx.doi.org/10.1155/2015/652026}{Adv. Math. Phys. 2015, 652026 (2015)}.

\bibitem{Mozo_QFT} L. \'Alvarez-Gaum\'e and M. \'{A}. V\'azquez-Mozo,
\emph{An Invitation to Quantum Field Theory} (Springer, Berlin, 2012).

\bibitem{Nielsen=000026Chuang} M. A. Nielsen and I. L. Chuang, \emph{Quantum
Computation and Quantum Information} (Cambridge University Press,
Cambridge, 2000).

\bibitem{Preskill} J. Preskill, \emph{Quantum Information and Computation},
\href{http:// theory.caltech.edu/people/preskill/ph229/}{http:// theory.caltech.edu/people/preskill/ph229/}.

\bibitem{Wilde} M. M. Wilde, \emph{Quantum Information Theory} (Cambridge
University Press, Cambridge, 2013).

\bibitem{Leuchs_RMP} M. D. Reid, P. D. Drummond, W. P. Bowen, E.
G. Cavalcanti, P. K. Lam, H. A. Bachor, U. L. Andersen, and G. Leuchs,
The Einstein-Podolsky-Rosen paradox: From concepts to applications,
\href{http://dx.doi.org/10.1103/RevModPhys.81.1727}{Rev. Mod. Phys. 81, 1727 (2009)}.

\bibitem{Horodecki_RMP} R. Horodecki, P. Horodecki, M. Horodecki,
and K. Horodecki, Quantum entanglement, \href{http://dx.doi.org/10.1103/RevModPhys.81.865}{Rev. Mod. Phys. 81, 865 (2009)}.

\bibitem{Jost_AMP} M. Li, S.-M. Fei, and X. Li-Jost, Quantum entanglement:
Separability, measure, fidelity of teleportation, and distillation,
\href{http://dx.doi.org/10.1155/2010/301072}{Adv. Math. Phys. 2010, 301072 (2010)}.

\bibitem{Lucas_RevD} L. C. C\'eleri, J. Maziero, and R. M. Serra,
Theoretical and experimental aspects of quantum discord and related
measures, \href{http://dx.doi.org/10.1142/S0219749911008374}{Int. J. Quantum Inf. 9, 1837 (2011)}.

\bibitem{Modi_RMP} K. Modi, A. Brodutch, H. Cable, T. Paterek, and
V. Vedral, The classical-quantum boundary for correlations: Discord
and related measures, \href{http://dx.doi.org/10.1103/RevModPhys.84.1655}{Rev. Mod. Phys. 84, 1655 (2012)}.

\bibitem{Lucas_PTRSA} D. O. Soares-Pinto, R. Auccaise, J. Maziero,
A. Gavini-Viana, R. M. Serra, and L. C. C\'eleri, On the quantumness
of correlations in nuclear magnetic resonance, \href{http://dx.doi.org/10.1098/rsta.2011.0364 }{Phil. Trans. R. Soc. A 370, 4821 (2012)}.

\bibitem{Brito_AMP} M. \'Avila, G. H. Sun, and A. L. Salas-Brito,
Scales of time where the quantum discord allows an efficient execution
of the DQC1 algorithm, \href{http://dx.doi.org/10.1155/2014/367905}{Adv. Math. Phys. 2014, 367905 (2014)}.

\bibitem{Streltsov_book} A. Streltsov, \emph{Quantum Correlations
Beyond Entanglement, and Their Role in Quantum Information Theory}
(SpringerBriefs in Physics, 2015).

\bibitem{Datta_DQC1} A. Datta, A. Shaji, and C. M. Caves, Quantum
discord and the power of one qubit, \href{http://dx.doi.org/10.1103/PhysRevLett.100.050502}{Phys. Rev. Lett. 100, 050502 (2008)}.

\bibitem{Winter_DSMerg} D. Cavalcanti, L. Aolita, S. Boixo, K. Modi,
M. Piani, and A. Winter, Operational interpretations of quantum discord,
\href{http://dx.doi.org/10.1088/PhysRevA.83.032324}{Phys. Rev. A 83, 032324 (2011)}.

\bibitem{Pirandola_DCryp} S. Pirandola, Quantum discord as a resource
for quantum cryptography, \href{http://dx.doi.org/10.1038/srep06956}{Sci. Rep. 4, 6956 (2014)}.

\bibitem{Retamal_DSDisc} L. Roa, J. C. Retamal, and M. Alid-Vaccarezza,
Dissonance is required for assisted optimal state discrimination,
\href{http://dx.doi.org/10.1103/PhysRevLett.107.080401}{Phys. Rev. Lett. 107, 080401 (2011).}

\bibitem{Piani_DDHid} M Piani, V Narasimhachar, and J Calsamiglia,
Quantumness of correlations, quantumness of ensembles and quantum
data hiding, \href{http://dx.doi.org/10.1088/1367-2630/16/11/113001}{New J. Phys. 16, 113001 (2014)}.

\bibitem{Vedral_RSP} B. Dakic, Y. O. Lipp, X. Ma, M. Ringbauer, S.
Kropatschek, S. Barz, T. Paterek, V. Vedral, A. Zeilinger, \v{C}. Brukner,
and P. Walther, Quantum discord as resource for remote state preparation,
\href{http://dx.doi.org/10.1038/NPHYS2377}{Nat. Phys. 8, 666 (2012)}.

\bibitem{Adesso_metro} D. Girolami, A. M. Souza, V. Giovannetti,
T. Tufarelli, J. G. Filgueiras, R. S. Sarthour, D. O. Soares-Pinto,
I. S. Oliveira, and G. Adesso, Quantum discord determines the interferometric
power of quantum states, \href{http://dx.doi.org/10.1103/PhysRevLett.112.210401}{Phys. Rev. Lett. 112, 210401 (2014)}.

\bibitem{Piani_DEComm} T. K. Chuan, J. Maillard, K. Modi, T. Paterek,
M. Paternostro, and M. Piani, Quantum discord bounds the amount of
distributed entanglement, \href{http://dx.doi.org/10.1103/PhysRevLett.109.070501}{Phys. Rev. Lett. 109, 070501 (2012)}.

\bibitem{Maziero_W1} J. Maziero and R. M. Serra, Classicality witness
for two-qubit states, \href{http://dx.doi.org/10.1142/S0219749912500281}{Int. J. Quantum Inf. 10, 1250028 (2012)}.

\bibitem{Maziero_W2} R. Auccaise, J. Maziero, L. C. Celeri, D. O.
Soares-Pinto, E. R. deAzevedo, T. J. Bonagamba, R. S. Sarthour, I.
S. Oliveira, and R. M. Serra, Experimentally witnessing the quantumness
of correlations, \href{http://dx.doi.org/10.1103/PhysRevLett.107.070501}{Phys. Rev. Lett. 107, 070501 (2011)}.

\bibitem{Sarandy_2sc_exp} F. M. Paula, I. A. Silva, J. D. Montealegre,
A. M. Souza, E. R. deAzevedo, R. S. Sarthour, A. Saguia, I. S. Oliveira,
D. O. Soares-Pinto, G. Adesso, and M. S. Sarandy, Observation of environment-induced
double sudden transitions in geometric quantum correlations, \href{http://dx.doi.org/10.1103/PhysRevLett.111.250401}{Phys. Rev. Lett. 111, 250401 (2013)}.

\bibitem{Adesso_OD} I. A. Silva, D. Girolami, R. Auccaise, R. S.
Sarthour, I. S. Oliveira, T. J. Bonagamba, E. R. deAzevedo, D. O.
Soares-Pinto, and G. Adesso, Measuring bipartite quantum correlations
of an unknown state, \href{http://dx.doi.org/10.1103/PhysRevLett.110.140501}{Phys. Rev. Lett. 110, 140501 (2013)}.

\bibitem{Maziero_W3} G. H. Aguilar, O. Jim\'enez Far\'ias, J. Maziero,
R. M. Serra, P. H. Souto Ribeiro, and S. P. Walborn, Experimental
estimate of a classicality witness via a single measurement, \href{http://dx.doi.org/10.1103/PhysRevLett.108.063601}{Phys. Rev. Lett. 108, 063601 (2012)}.

\bibitem{Bose_AMP} A. Bayat and S. Bose, Entanglement transfer through
an antiferromagnetic spin chain, \href{http://dx.doi.org/10.1155/2010/127182}{Adv. Math. Phys. 2010, 127182 (2010)}.

\bibitem{LoFranco1} M. Cianciaruso, T. R. Bromley, W. Roga, R. Lo
Franco, and G. Adesso, Universal freezing of quantum correlations
within the geometric approach, \href{http://dx.doi.org/10.1038/srep10177}{Sci. Rep. 5, 10177 (2015)}.

\bibitem{LoFranco2} T. R. Bromley, M. Cianciaruso, R. Lo Franco,
and G. Adesso, Unifying approach to the quantification of bipartite
correlations by Bures distance, \href{http://dx.doi.org/10.1088/1751-8113/47/40/405302}{J. Phys. A: Math. Theor. 47, 405302 (2014)}.

\bibitem{LoFranco3} B. Aaronson, R. Lo Franco, G. Compagno, and G.
Adesso, Hierarchy and dynamics of trace distance correlations, \href{http://dx.doi.org/10.1088/1367-2630/15/9/093022}{New J. Phys. 15, 093022 (2013)}.

\bibitem{LoFranco4} B. Aaronson, R. Lo Franco, and G. Adesso, Comparative
investigation of the freezing phenomena for quantum correlations under
nondissipative decoherence, \href{http://dx.doi.org/10.1103/PhysRevA.88.012120}{Phys. Rev. A 88, 012120 (2013)}.

\bibitem{LoFranco5} B. Bellomo, G. L. Giorgi, F. Galve, R. Lo Franco,
G. Compagno, and R. Zambrini, Unified view of correlations using the
square norm distance, \href{http://dx.doi.org/10.1103/PhysRevA.85.032104}{Phys. Rev. A 85, 032104 (2012)}.

\bibitem{LoFranco6} B. Bellomo, R. Lo Franco, and G. Compagno, Dynamics
of geometric and entropic quantifiers of correlations in open quantum
systems, \href{http://dx.doi.org/10.1103/PhysRevA.86.012312}{Phys. Rev. A 86, 012312 (2012)}.

\bibitem{LoFranco7} R. Lo Franco, B. Bellomo, E. Andersson, and G.
Compagno, Revival of quantum correlations without system-environment
back-action, \href{http://dx.doi.org/10.1103/PhysRevA.85.032318}{Phys. Rev. A 85, 032318 (2012)}.

\bibitem{LoFranco8} J.-S. Xu, K. Sun, C.-F. Li, X.-Y. Xu, G.-C. Guo,
E. Andersson, R. Lo Franco, and G. Compagno, Experimental recovery
of quantum correlations in absence of system-environment back-action,
\href{http://dx.doi.org/10.1038/ncomms3851}{Nat. Commun. 4, 2851 (2013)}.

\bibitem{LoFranco9} R. Lo Franco, B. Bellomo, S. Maniscalco, and
G. Compagno, Dynamics of quantum correlations in two-qubit systems
within non-Markovian environments, \href{http://dx.doi.org/10.1142/S0217979213450537}{Int. J. Mod. Phys. B 27, 1345053 (2013)}.

\bibitem{Maziero_PTr} J. Maziero, Computing partial traces and reduced
density matrices, \href{http://arxiv.org/abs/1601.07458}{arXiv:1601.07458}.

\bibitem{Petruccione_param} E. Br\"{u}ning, H. M\"{a}kel\"{a}, A.
Messina, and F. Petruccione, Parametrizations of density matrices,
\href{http://dx.doi.org/10.1080/09500340.2011.632097}{J. Mod. Opt. 59, 1 (2012)}.

\bibitem{Bertlmann} R. A. Bertlmann and P. Krammer, Bloch vectors
for qudits, \href{http://dx.doi.org/10.1088/1751-8113/41/23/235303}{J. Phys. A: Math. Theor. 41, 235303 (2008)}.

\bibitem{Maziero_FICT} J. Maziero, Fortran code for generating random
probability vectors, unitaries, and quantum states, \href{http://dx.doi.org/10.3389/fict.2016.00004}{Front. ICT 3, 4 (2016)}.

\bibitem{Huang_DiscCC} Y. Huang, Computing quantum discord is NP-complete,
\href{http://dx.doi.org/10.1088/1367-2630/16/3/033027}{New J. Phys. 16, 033027 (2014)}.

\bibitem{Sen_DiscCC} T. Chanda, T. Das, D. Sadhukhan, A. K. Pal,
A. Sen De, and U. Sen, Reducing computational complexity of quantum
correlations, \href{http://dx.doi.org/10.1103/PhysRevA.92.062301}{Phys. Rev. A 92, 062301 (2015)}.

\bibitem{Luo_DiscAn1} S. Luo, Quantum discord for two-qubit systems,
\href{http://dx.doi.org/10.1103/PhysRevA.77.042303}{Phys. Rev. A 77, 042303 (2008)}.

\bibitem{Luo_DiscAn2} S. Luo and Q. Zhang, Observable correlations
in two-qubit states, \href{http://dx.doi.org/10.1007/s10955-009-9779-0}{J. Stat. Phys. 136, 165 (2009)}.

\bibitem{Alber_DiscAnal_XS} M. Ali, A. R. P. Rau, and G. Alber, Quantum
discord for two-qubit X states, \href{http://dx.doi.org/10.1103/PhysRevA.81.042105}{Phys. Rev. A 81, 042105 (2010)}.

\bibitem{Maziero_DiscAn} J. Maziero, L. C. Celeri, and R. M. Serra,
Symmetry aspects of quantum discord, \href{http://arxiv.org/abs/1004.2082}{arXiv:1004.2082}.

\bibitem{Adesso_DAnal} D. Girolami and G. Adesso, Quantum discord
for general two-qubit states: Analytical progress, \href{http://dx.doi.org/10.1103/PhysRevA.83.052108}{Phys. Rev. A 83, 052108 (2011)}.

\bibitem{Oh_DAnal} Q. Chen, C. Zhang, S. Yu, X. X. Yi, and C. H.
Oh, Quantum discord of two-qubit X states, \href{http://dx.doi.org/10.1103/PhysRevA.84.042313}{Phys. Rev. A 84, 042313 (2011)}.

\bibitem{Joag_DhsAnal} A. S. M. Hassan, B. Lari, and P. S. Joag,
Tight lower bound to the geometric measure of quantum discord, \href{http://dx.doi.org/10.1103/PhysRevA.85.024302}{Phys. Rev. A 85, 024302 (2012)}.

\bibitem{Giovannetti_AnalDtr} F. Ciccarello, T. Tufarelli, and V.
Giovannetti, Toward computability of trace distance discord, \href{http://dx.doi.org/10.1088/1367-2630/16/1/013038}{New J. Phys. 16, 013038 (2014)}.

\bibitem{Orzag_DAnal} D. Spehner and M. Orszag, Geometric quantum
discord with Bures distance: the qubit case, \href{http://dx.doi.org/10.1088/1751-8113/47/3/035302}{J. Phys. A: Math. Theor. 47, 035302 (2014)}.

\bibitem{Zahir_DAnal} M. A. Jafarizadeh, N. Karimi, and H. Zahir,
Quantum discord for generalized Bloch sphere states, \href{http://dx.doi.org/10.1140/epjd.e2014-40677-6}{Eur. Phys. J. D 68, 136 (2014)}.

\bibitem{Seddik_DAnal} M. Daoud, R. Ahl Laamara, and S. Seddik, Hilbert-Schmidt
measure of pairwise quantum discord for three-qubit X states, \href{http://dx.doi.org/10.1016/S0034-4877(15)30030-6}{Rep. Math. Phys. 76, 207 (2015)}.

\bibitem{Bordone_DAnal} A. Beggi, F. Buscemi, and P. Bordone, Analytical
expression of genuine tripartite quantum discord for symmetrical X-states,
\href{http://dx.doi.org/10.1007/s11128-014-0882-z}{Quantum Inf. Process. 14, 573 (2015)}.

\bibitem{Fan_DAnal} Y.-K. Wang, N. Jing, S.-M. Fei, Z.-X. Wang, J.-P.
Cao, and H. Fan, One-way deficit of two-qubit X states, \href{http://dx.doi.org/10.1007/s11128-015-1005-1}{Quantum Inf. Process. 14, 2487 (2015)}.

\bibitem{Fei_DAnal} Z. Ma, Z. Chen, F. F. Fanchini, and S.-M. Fei,
Quantum discord for $d\otimes2$ systems. \href{http://dx.doi.org/10.1038/srep10262}{Sci. Rep. 5, 10262 (2015)}.

\bibitem{Wang_DAnal} N. Jing, X. Zhang, and Y.-K. Wang, Comment on
``One-way deficit of two qubit X states'', \href{http://dx.doi.org/10.1007/s11128-015-1132-8}{Quantum Inf. Process. 14, 4511 (2015)}.

\bibitem{Zhang_DAnal} G. Li, Y. Liu, H. Tang, X. Yin, and Z. Zhang,
Analytic expression of quantum correlations in qutrit Werner states
undergoing local and nonlocal unitary operations, \href{http://dx.doi.org/10.1007/s11128-014-0888-6}{Quantum Inf. Process. 14, 559 (2015)}.

\bibitem{Sarandy_Dtr_Anal} P. C. Obando, F. M. Paula, and M. S. Sarandy,
Trace-distance correlations for X states and the emergence of the
pointer basis in Markovian and non-Markovian regimes, \href{http://dx.doi.org/10.1103/PhysRevA.92.032307}{Phys. Rev. A 92, 032307 (2015)}.

\bibitem{Huang_2} Y. Huang, Quantum discord for two-qubit X states:
Analytical formula with very small worst-case error, \href{http://dx.doi.org/10.1103/PhysRevA.88.014302}{Phys. Rev. A 88, 014302 (2013)}.

\bibitem{Modi_DCriteria} A. Brodutch and K. Modi, Criteria for measures
of quantum correlations, \href{http://arxiv.org/abs/1108.3649}{Quantum Inf. Comp. 12, 0721 (2012)}.

\bibitem{Brukner_HSD} B. Daki\'{c}, V. Vedral, and \v{C}. Brukner,
Necessary and sufficient condition for nonzero quantum discord, \href{http://dx.doi.org/10.1103/PhysRevLett.105.190502}{Phys. Rev. Lett. 105, 190502 (2010)}.

\bibitem{Luo_HSD} S. Luo and S. Fu, Geometric measure of quantum
discord, \href{http://dx.doi.org/10.1103/PhysRevA.82.034302}{Phys. Rev. A 82, 034302 (2010)}.

\bibitem{Akhtarshenas1} S. J. Akhtarshenas, H. Mohammadi, S. Karimi,
and Z. Azmi, Computable measure of quantum correlation, \href{http://dx.doi.org/10.1007/s11128-014-0839-2}{Quantum Inf. Process. 14, 247 (2015)}.\end{thebibliography}
\end{document}